\documentclass[prl,twocolumn,superscriptaddress]{revtex4-1}
\usepackage{amsmath, amssymb,graphicx,color,bm,epstopdf}
\usepackage[dvipsnames]{xcolor}
\usepackage{bbold}
\usepackage{braket}
\usepackage{comment}
\usepackage[unicode=true,colorlinks=true,citecolor=blue,urlcolor=blue]{hyperref}
\usepackage{graphicx}

\newcommand{\beml}{\begin{subequations}}
\newcommand{\eml}{\end{subequations}}
\newcommand{\beq}{\begin{eqnarray}}
\newcommand{\eq}{\end{eqnarray}}
\newcommand{\ba}{\begin{array}}
\newcommand{\ea}{\end{array}}
\newcommand{\bpm}{\begin{pmatrix}}
\newcommand{\epm}{\end{pmatrix}}
\newcommand{\bc}{\begin{cases}}
\newcommand{\ec}{\end{cases}}

\renewcommand {\Im}{\mathop\mathrm{Im}\nolimits}

\renewcommand {\phi}{{\varphi}}
\newcommand {\rmi}{{\rm i}}
\newcommand {\PT}{{$\mathcal{PT}$}{}}
\newcommand {\rmd}{{\rm d}}

\newcommand {\e}{{\rm e}}

\usepackage[utf8]{inputenc}
\begin{document}
\title{Waveguide quantum optomechanics: parity-time phase transitions in ultrastrong coupling regime}
\author{Ivan Iorsh}
\affiliation{Department of Physics and Technology, ITMO University, St. Petersburg, 197101, Russia}
\author{Alexander Poshakinskiy}
\affiliation{Ioffe Institute, 194021, St. Petersburg, Russia}
\author{Alexander Poddubny}
\affiliation{Ioffe Institute, 194021, St. Petersburg, Russia}
\affiliation{Department of Physics and Technology, ITMO University, St. Petersburg, 197101, Russia}
\begin{abstract}
We develop a  rigorous theoretical framework for interaction-induced phenomena in the  waveguide quantum electrodynamics (QED) driven by   mechanical oscillations of the qubits. Specifically, we predict  that  the simplest set-up of two qubits,  harmonically trapped  over an optical waveguide, enables the ultrastrong coupling regime of the quantum optomechanical interaction. Moreover, the combination of the inherent open nature of the system and the strong optomechanical coupling leads to  emerging  parity-time (\PT) symmetry, quite unexpected  for a  purely quantum system without artificially engineered gain and loss.
The $\mathcal{PT}$ phase transition drives   long-living subradiant states, observable in the state-of-the-art waveguide QED setups. 
\end{abstract}

\maketitle
\textit{Introduction.}
Waveguide quantum electrodynamics~\cite{Roy2017,KimbleRMP2018}, studying the interaction of  quantum emitters with propagating photons,   experiences now a  rapid progress driven by novel quantum technologies. Man-made arrays of cold atoms~\cite{Corzo2019} and superconducting qubits~\cite{vanLoo2013,Mirhosseini2019} allow one to explore novel topological physics~\cite{kim2020quantum,Chang2020}, collective super-radiance and sub-radiance ~\cite{Ke2019,kornovan2019extremely,Zhang2019arXiv} with unprecedented  tunability and precision and can be used to build future quantum networks~\cite{kimble2008quantum}. 
However, the photon-qubit interaction in the waveguides is intrinsically weaker  than in the cavity setup, since light is delocalized in space. It is not clear whether the  fundamentally interesting regime of quantum phase transitions in ultra-strong coupling regime~\cite{FriskKockum2019} can  be demonstrated in waveguide QED.

In this Letter  we predict that mechanical motion of  the qubits coupled to the waveguide opens up a new route to ultrastrong coupling at the quantum level for experimentally accessible parameters. Despite recent theoretical advances~\cite{manzoni2017,Lesanovsky2020,snchezburillo2020theory} and experimental observations of optomechanical nonreciprocity with  superconducting qubits~\cite{Peterson2017},  the field of waveguide quantum optomechanics still remains practically unchartered.   Here we develop a rigorous theoretical framework to describe light-mediated coupling between mechanical oscillations in the array of  qubits,  harmonically trapped  over an optical waveguide. Our calculations reveal an unexpectedly rich behavior  of  coupled light, qubits and vibrations. Even in the simplest case of just two qubits
\begin{figure}[!b]
    \centering
    \includegraphics[width=0.9\columnwidth]{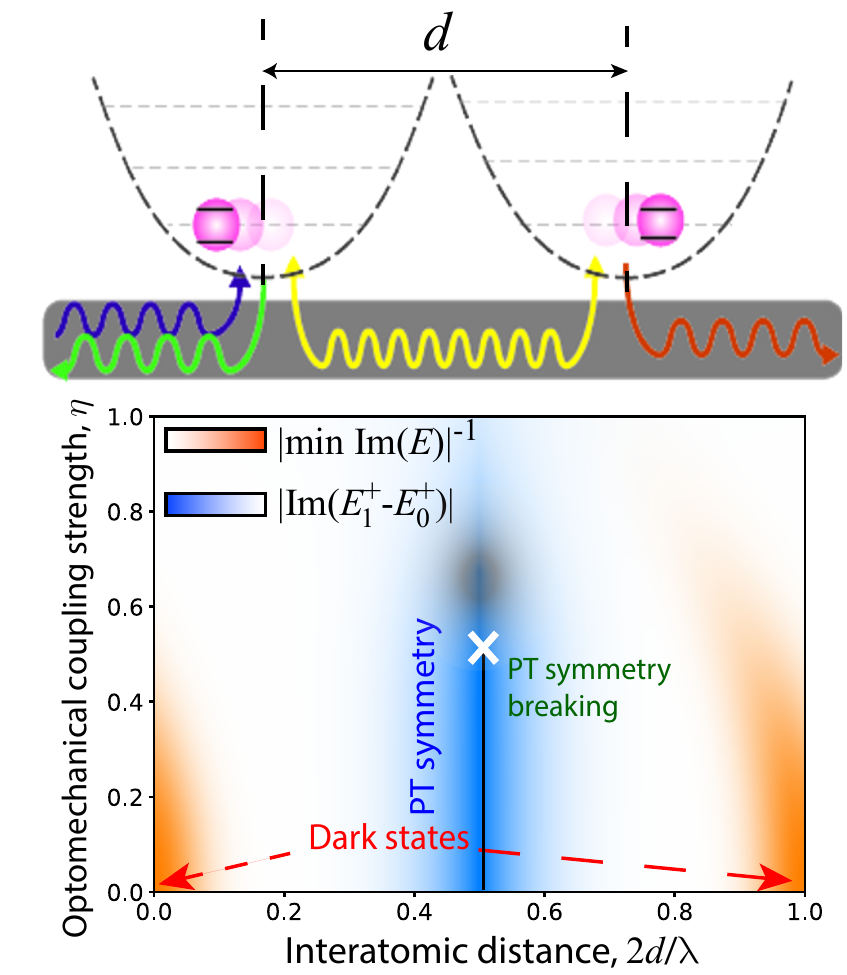}
    \caption{ (Upper panel) Geometry of the problem: two oscillating qubits, trapped in the parabolic potentials, and coupled to the waveguide. (Lower panel) Phase diagram showing by color the radiative decay rates $-\Im E_0^{\pm}$ of the (anti)symmetric superposition of the photonic qubit excitations  in the lowest vibrational state as function of interatomic distance and optomechanical coupling strength calculated from Hamiltonian~\eqref{H_d}. Red color indicates small decay rates, with color saturation proportional to the lifetime of most long-lived state $-1/\text{min}(\Im E_0^{+},\Im E_0^{-})$. Blue color encodes the difference between the decay rates of the two lowest vibrational states $|\Im (E_1^+ - E_0^+)|$ with higher saturation corresponding to smaller difference realized in the vicinity of unbroken \PT{} symmetry (solid line). }
    \label{fig1}
\end{figure}
  the system is described by a celebrated quantum Rabi model~\cite{Casanova2018}, which, in stark contrast to the usual situation, is inherently non-Hermitian due to the possibility of radiative decay into the waveguide.  We examine  the  phase diagram  depending on the strength of light-matter and optomechanical couplings and find the phases   very sensitive to the inter-qubit spacings, controlling the radiative decay. Namely,  when the spacing is equal to the quarter of light wavelength  at the qubit resonance $\lambda/4$,  the system demonstrates an emergence of parity-time (\PT) symmetry~\cite{Feng2017}. The \PT{} symmetry breaking  drives  formation of coupled modes of light and vibrations with suppressed radiative decay, that manifest themselves as  narrow resonances of single-photon transmission. A very different  scenario is observed in case of  Dicke superradiance, when the spacing is an integer multiple of $\lambda/2$. When the qubits are still, the spectrum has one subradiant and one subradiant mode, while  an increase of optomechanical coupling brightens the  subradiant mode and enables high-quality resonances in optical spectra. Both of these scenarios are summarized at the lower panel of Fig.~\ref{fig1}. Our results demonstrate that the optomechanical coupling opens a new degree of freedom to engineer radiative lifetimes, spatial distributions  and quantum correlations in waveguide quantum electrodynamics.  

\textit{Optomechanical Quantum Rabi Model.}
We consider an array of qubits schematically shown in Fig.~\ref{fig1}. Each  qubit is trapped in individual harmonic optical trap which is placed in the vicinity of a single-mode optical fiber. The qubit  can absorb or emit a waveguide photon, and the radiative relaxation to the far field outside the waveguide is suppressed. The Hamiltonian of the system can be written as
\begin{align}
\hat{H}=\sum_k \omega_k c_k^{\dagger}c_k +\sum_j \omega_x\sigma_j^{+}\sigma_j+\sum_j\Omega a_j^{\dagger}a_j +\hat{H}_{\rm int},
\end{align}
where $\omega_k=v|k|$ is the dispersion of the waveguide modes, which is assumed to be linear, $\omega_x$ is the qubit resonance frequency, and $\Omega$ is the optical trap phonon energy. The interaction Hamiltonian is given by
\begin{align}
\hat{H}_{\rm int}=g\sum_{k,j}\left[\sigma_j^{\dagger}c_k^{\vphantom{\dagger}}\e^{ik[z_j+u_0(a_j+a_j^{\dagger})]}+\mathrm{H.c.}\right],
\end{align}
where $g$ is the Rabi splitting, $z_j$ is the position of the $j$-th optical trap, and $u_0=\sqrt{\hbar/(2M\Omega)}$ is the quantum of the mechanical motion, where $M$ is the mass of the qubit.

We  integrate out the waveguide degrees of freedom~\cite{Shi2011}
to obtain the effective Hamiltonian up to the second order of the qubit-photon coupling $g$:
\begin{align}
&\hat{H}_{\rm eff}=\sum_j \omega_x\sigma_j^{+}\sigma_j+\sum_j\Omega a_j^{\dagger}a_j\nonumber\\-&i\Gamma_0\sum_{jl}\sigma_j^{+}\sigma_le^{iq|z_j-z_l|}e^{iqu_0 \,\text{sign}(z_j-z_l)\,(a_j+a_j^{\dagger}-a_l-a_l^{\dagger})},  \label{H_eff}
\end{align}
where $\Gamma_0=g^2/v$ is the radiative decay rate of a single qubit and we used the Markov approximation neglecting the frequency dispersion in the phase factor, $k\approx q \equiv \omega_x/v$. Hamiltonian~\eqref{H_eff} can be also obtained  from the conventional waveguide-QED  Hamiltonian~\cite{Molmer2019,Ke2019} where the coordinates of static qubits $z_j$ are replaced with the corresponding dynamical value $z_j+ u_0(a_j+a_j^{\dagger})$ and $u_0 \lesssim |z_j-z_l|$ is supposed. 

In this work we restrict ourselves to a specific case of only two qubits and only a single photonic  excitation in the array. Then, the qubit subspace of the full Hilbert space is a span of only two states $|L\rangle$ and $|R\rangle$ corresponding to the left/right qubit being in the excited state. 
We introduce a new rotated basis comprising a symmetric and antisymmetric superposition of the excitations at the individual qubits: $|\pm\rangle =\frac{1}{\sqrt{2}}(|L\rangle\pm |R\rangle)$. Moreover, we perform the unitary transformation of the phononic degrees of freedom: $\hat{x}_{s(d)}=(\hat{x}_2\pm \hat{x}_1)/\sqrt{2}$, where $\hat{x}_i=a_i+a_i^{\dagger}$. In the new basis the effective Hamiltonian is written as
\begin{align}
\hat{H}_{\rm eff}=\Omega a_s^{\dagger}a_s^{\vphantom{\dagger}} +\Omega a_d^{\dagger}a_d^{\vphantom{\dagger}} -i\Gamma_0\left(1+\sigma_ze^{i\phi} e^{i\eta \hat{x}_d}\right), \label{H_efffull}
\end{align}
where $\phi=q|z_1-z_2|$ and $\eta=2qu_0$. We note, that the symmetric vibrations are decoupled from the rest part of the Hamiltonian since they do not affect the distance between the qubits. Thus, the centre of mass of the qubits oscillates freely and the number of $s$ phonons is a good quantum number. Moreover, the $|+\rangle$ and $|-\rangle$ states are also decoupled and described by the separate Hamiltonians:
\begin{align}\label{H_d0}
\hat{H}_{\rm eff}^{\pm}=\Omega a_d^{\dagger}a_d^{\vphantom{\dag}} -i\Gamma_0\left(1\pm e^{i\phi} e^{i\eta \hat{x}_d}\right).
\end{align}
Namely, when $\eta=0,\phi=\pi$ the antisymmetric state corresponds to a dark, subradiant state, and the symmetric one to a superradiant state.

The spectrum of the obtained Hamiltonians can be found most conveniently by using the real-space representation of the mechanical motion. Namely, introducing the  coordinate $x_d$ measured in units of $\sqrt{2}u_0$ we obtain the following representation of the effective Hamiltonian
\begin{align}
\hat H_{\rm eff}^{\pm} =\frac{\Omega}{2}\left(-\frac{\rmd^2}{\rmd x_d^2}+x_d^2-1\right)-i\Gamma_0\mp i\Gamma_0e^{i\phi}e^{i\eta x_d}. ~\label{H_d}
\end{align}
The obtained Hamiltonian allows for a simple numerical solution with the finite difference schemes. 

Before we proceed further, it is important to discuss the range of the experimentally accessible parameters of the model. The relative strength of the quantum optomechanical coupling is defined by the ratio of the length scale of the mechanical atomic movement and the wavelength of the photon in the wavequide, $\eta=4\pi u_0/\lambda$. The length scale of the atomic movement can be roughly estimated via the de Broglie wavelength $u_0<\hbar/p_{\rm th}$, where the thermal momentum $p_{\rm th}=\sqrt{3Mk_BT}$. For the lithium atoms and the resonant wavelength approximately 700 nm the value of $\eta=1$ is achieved at $T=640$ nK, which is a  temperature which has been achieved in recent cold atom experiments (see the review~\cite{anglin2002bose} and references within). The corresponding phonon energy is then approximately $2.4$ kHz. The radiative decay rate $\Gamma_0$ can be flexibly tuned in a wide range of frequencies from zero to the GHz. Thus, the range of $\Gamma_0/\Omega,\eta \sim 1$ is achievable within the modern experimental techniques.

Importantly,  the Hamiltonians \eqref{H_d0} and~\eqref{H_d} have strong similarity with the  quantum Rabi model~\cite{Casanova2018}. Indeed, the equidistant harmonic oscillator eigenstates with energy separation $\Omega$ play the role of the photon number states in the Rabi model. The third term in Eq.~\eqref{H_d} mixes the different eigenstates similar to the atom-light interaction term in the Rabi model, and the decay rate  $\Gamma_0$ plays the role of the effective Rabi frequency. Since $\Omega$ is small, the ratio $\Gamma_0/\Omega$ can  be arbitrary large and it is possible to access the so-called ultrastrong coupling regime~\cite{kockum2019ultrastrong}. In what follows, the energy will be normalized to the phonon energy $\Omega$.
\begin{figure}[!h]
    \centering
    \includegraphics[width=1.0\columnwidth]{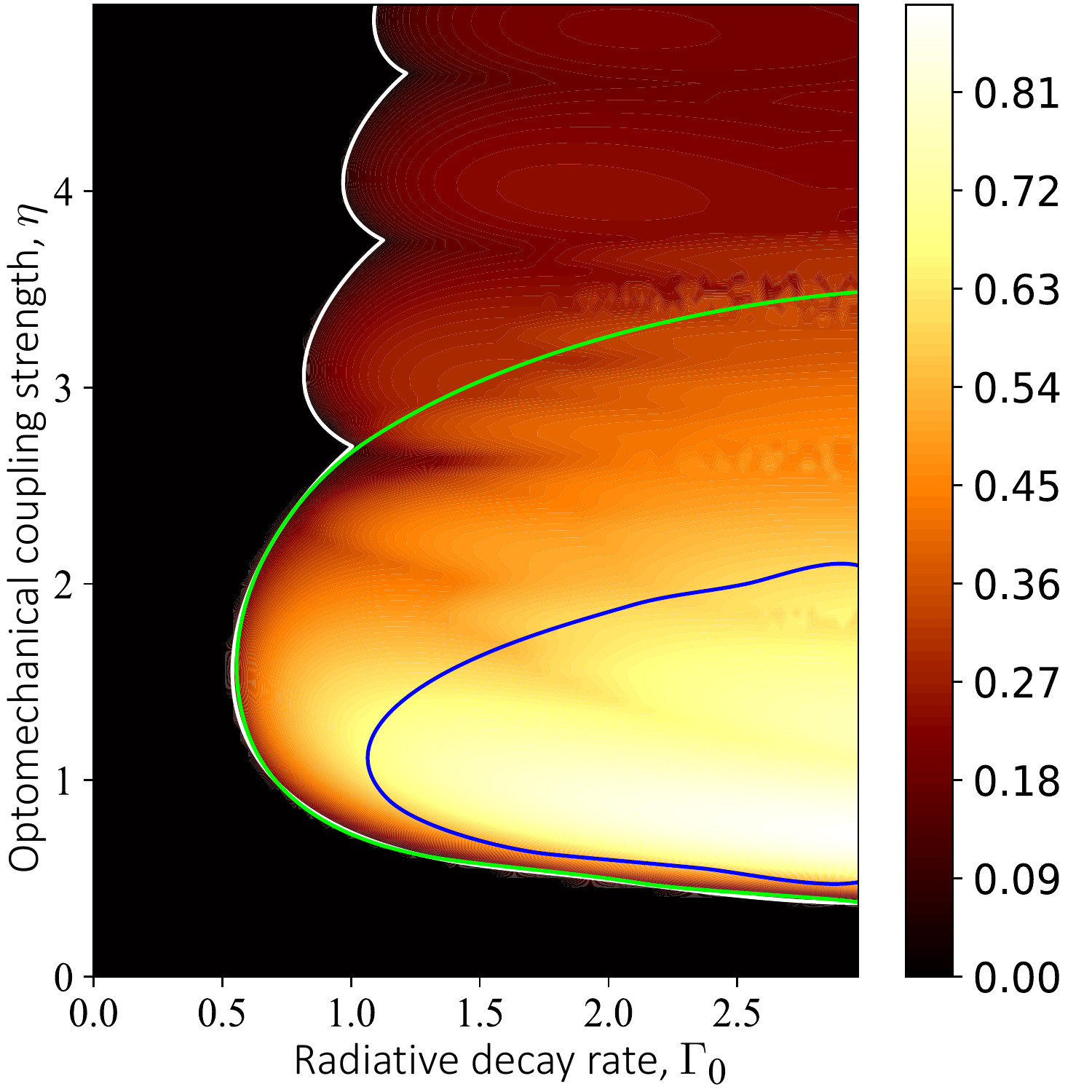}
    \caption{Optomechanical phase diagram for two $\lambda/4$-spaced qubits . The green and blue lines show the  \PT{}-phase boundaries for the first and second pair of the lowest symmetric eigenstates, respectively. The white line separates of the \PT{}-preserved and \PT{}-broken phase. The colorbar corresponds to $\mathrm{max}(\Im E)/\Gamma_0$.}
    \label{fig:PD}
\end{figure}
In stark contrast to the usual Rabi model~\cite{Casanova2018}, the Hamiltonians \eqref{H_d0} and~\eqref{H_d} are inherently non-Hermitian since the photons can escape into the waveguide. The strength of the radiative decay is controlled by the  distance between the qubits determining the phase $\varphi$.
We now set $\phi=\pi/2$, which means that the qubits are spaced by the quarter of light wavelength at the resonance frequency. In this case,  up to the constant  decay $\Gamma_0$, the Hamiltonians in Eq.~\eqref{H_d} become parity-time ($\mathcal{PT}$) symmetric. Indeed, the Hamiltonians $H^{\rm \pm}_{\rm eff}+i\Gamma_0$ are invariant under the simultaneous action of the parity ($x\rightarrow -x$) and time reversal (complex conjugation) operators.

\textit{$\mathcal{PT}$ phase transition}.
The parity-time symmetric Hamiltonians can either have real eigenvalues (unbroken-\PT{} phase) or pairs of complex conjugated eigenvalues  (broken-\PT{} phase)~\cite{Feng2017}.  The \PT{} phase transition occurs at a certain value of coupling strength. 
 In Fig.~\ref{fig:PD} we plot the numerically calculated \PT{} phase diagram for the Hamiltonians~\eqref{H_d} depending on the strength of the light-matter coupling $\Gamma_0$ and the optomechanical coupling strength $\eta$. The white solid line separates the \PT{}-preserved phase (black area) where all the eigenvalues are real and the \PT-broken phase, when at least one of the eigenenergies becomes complex. The colorbar shows the ratio of the largest imaginary part to $\Gamma_0$.  The green and blue lines correspond to the \PT{}-phase boundary for the first and second pair of the lowest symmetric eigenstates, respectively. The phase diagram demonstrates an existence of  \PT{}-unbroken phase for arbitrary large $\eta$ when $\Gamma_0<1/2$. As $\Gamma_0$ increases, the threshold $\eta$ for the \PT{}-phase transition decreases slowly.
\begin{figure}[!t]
    \centering
    \includegraphics[width=1.0\columnwidth]{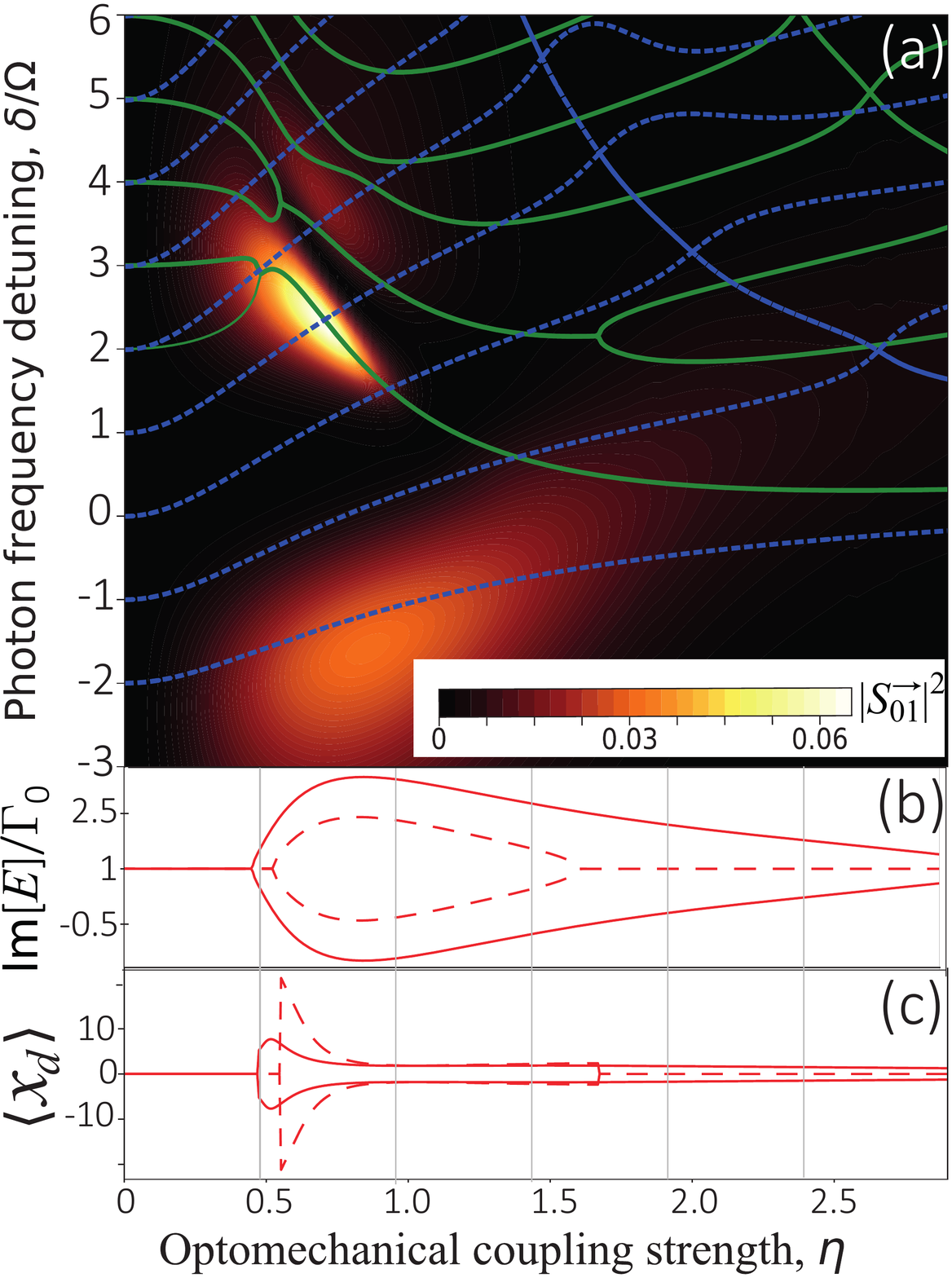}
    \caption{(a) Evolution of the real parts of the eigenergies of $H_{\rm eff}^{+}$ (solid green lines) and $H_{\rm eff}^{-}$ (dashed blue lines). The color map shows the spectrum of the transmission coefficient with absorption of a single $d$ phonon $|S^{\rightarrow}_{0,1}(\omega)|^2$ expressed via Eq.~\eqref{r_full}. (b,c) Evolution of the imaginary part of eigenergy (b) and matrix element $\langle x_d\rangle$ (c) for the first four eigenstates of $H_{\rm eff}^{+}$ vs optomechanical coupling $\eta$. Calculation has been performed for $\Gamma_0/\Omega=2.$  } 
    \label{fig:T01}
\end{figure}

We will now analyze the \PT{} phase transition in more detail. To this end, we  fix  the light-matter coupling strength to $\Gamma_0=2\Omega$ and plot in Fig.~\ref{fig:T01}(a) the real parts of the eigenergies $E^{\pm}$ with green and blue lines, respectively. The spectrum of the imaginary parts of  the four lowest eigenenergies of $H_{\rm eff}^{+}$ is shown in Fig.~\ref{fig:T01}(b). Naturally, at $\eta=0$ the spectrum is real and  it is essentially a pair  of the harmonic oscillator ladders with the energies $\pm \Gamma_0  +n\Omega$, with $n=0,1,2\ldots$. As $\eta$ increases, the energies of the first two pairs of eigenstates of $H_{\rm eff}^{+}$ (solid green lines) approach each other, and collapse at the corresponding threshold values of $\eta$ close to $0.5$.  This is the point of  \PT{} phase transition, after which the energies acquire imaginary parts and form complex conjugate pairs, see Fig.~\ref{fig:T01}(b). At even higher values $\eta$, the second phase transition occurs, where  the imaginary parts of the eigenenergies collapse. Consequently, there exists a certain critical value of $\eta$, where the maximum deviation of the eigenenergy imaginary part from $-\Gamma_0$ is achieved. At this point, one of the states, is characterized by the suppressed decay rate. We thus observe the optomechanically induced darkening of the qubit state.

{The breaking of the $\mathcal{PT}$ symmetry can be directly observed in the dependence of the average inter-qubit displacement $\langle x_d \rangle=\int dx_d \psi_{n,+}^{*}x_d\psi_{n,+}$ on $\eta$ shown in Fig.~\ref{fig:T01}(c). 
For small $\eta$, when the eigenstates are $\mathcal{PT}$-symmetric, the average value of the $\mathcal{PT}$-odd quantity $\langle x_d \rangle$ must be zero, and thus the qubits are placed at the equilibrium positions. At the $\mathcal{PT}$ symmetry breaking point, $\langle{x_d}\rangle$ changes discontinuously, and a pair of eigenstates connected by $\mathcal{PT}$  symmetry operation acquire opposite nonzero values of $\langle{x_d}\rangle$. The qubit pair becomes 'stretched' for the darker and 'squeezed' for brighter state. The effect of squeezing or stretching has been considered in the qubit chains in the different regime  when both coordinates and qubit polarizations were treated classically~\cite{Chang2013}}

We will now demonstrate that  the  subradiant modes, resulting from quantum optomechanical darkening,  are manifested as sharp resonances  in the spectra of single photon scattering. To calculate the latter, we exploit the Green's function, described in the Supplementary Material in detail.  The coefficients of single photon forward (backward) scattering accompanied by emission of $N_s$ $s$-phonons and  $N_d$ $d$-phonons read
%
%
\begin{align}
&S^{\rightarrow(\hookleftarrow)}_{N_s,N_d}(\omega)=- i\Gamma_0\sum_{N_s'=0}^{\infty}\sum_{N_d'=0}^{\infty}\sum_{p=\pm}p^{\frac{1\mp1}2}\int 
dx_sdx'_sdx_{d}dx'_d \nonumber\\
&\times\frac{e^{i\frac{\eta}{2}(x_s \mp x_s')}\left[\cos(\frac{\eta}{2}(x_d'-x_d))+ p \cos(\phi+\frac{\eta}{2}(x_d+x_d'))\right]}{\omega-\omega_x-E_{N_d'}^{\pm}-N_s'\Omega} \nonumber\\
&\times \phi_{N_s}(x_s)\phi_{N'_s}(x_s)\phi_{N'_s}({x'_s})\phi_0(x'_s)  \nonumber\\
&\times\phi_{N_d}(x_d)\psi^{\pm}_{N'_d}(x_d)\psi^{\pm}_{N'_d}(x'_d)\phi_0(x'_d)
, \label{r_full}
\end{align}
where $\phi_N(x)$ is the eigenfunction of the free harmonic oscillator corresponding to $N$ phonons and summation is taken over all possible intermediate states, characterised by the photonic mode parity $p$ and the number of  mechanical excitation, $N_s'$ and $N_d'$. The energy of the intermediate states is $E_{N_d'}^{\pm}+N_s'\Omega$, where $E_{N_d'}^{\pm}$ is the eigenenergy of $H_{\rm eff}^{\pm}
$ and $\psi^{\pm}_{N_d'}(x)$ is the corresponding wave function normalized according to $\int [\psi^{\pm}_{N_d'}(x)]^2\,dx = 1$. 
Note that that while the phonon reflection probability is simply $|S^{\hookleftarrow}_{N_s,N_d}(\omega)|^2$, that of photon transmission reads $|\delta_{N_s,0}\delta_{N_d,0}+S^{\rightarrow}_{N_s,N_d}(\omega)|^2$. 
\begin{figure}[!t]
    \centering
    \includegraphics[width=1.0\columnwidth]{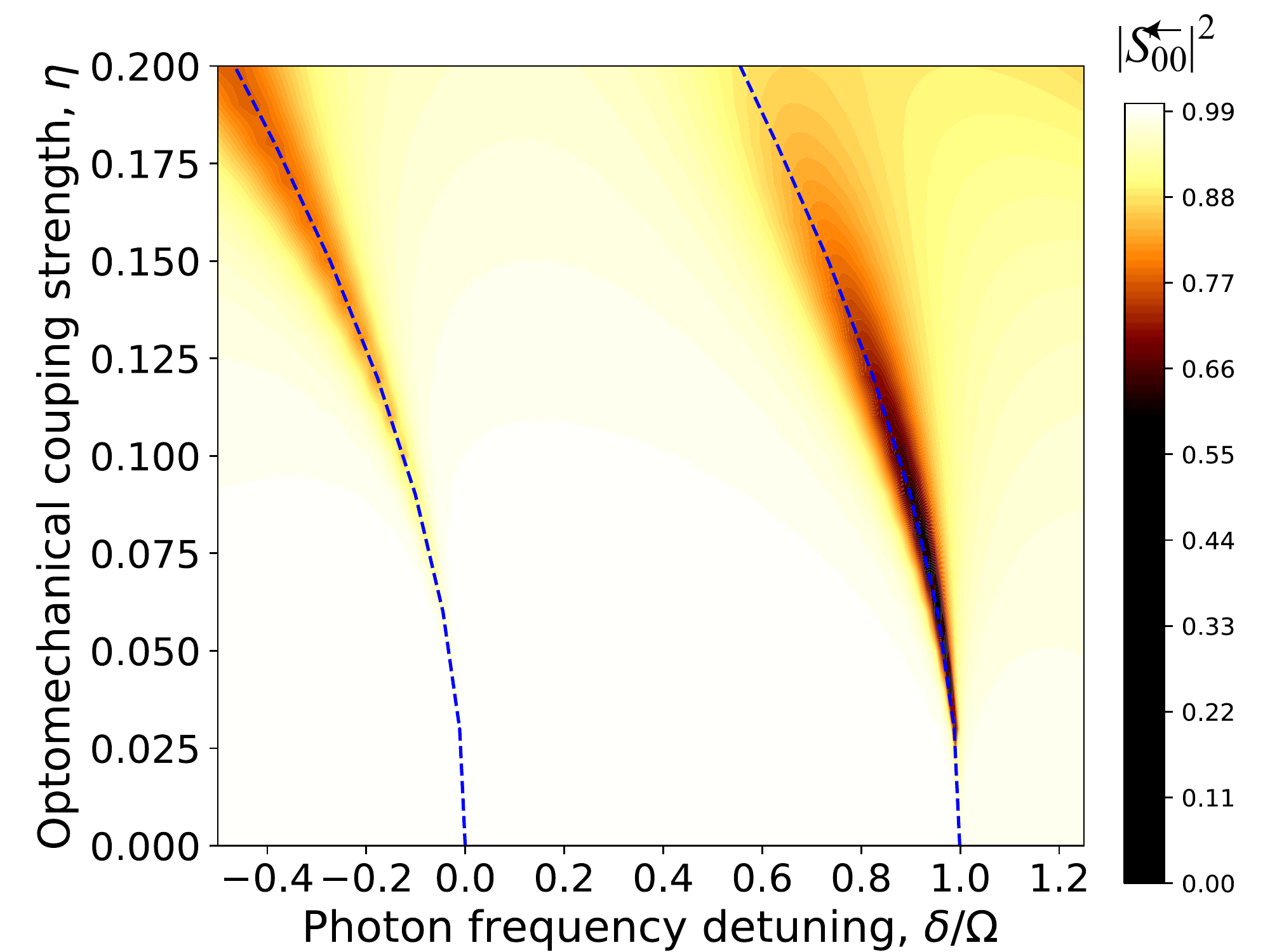}
    \caption{ Evolution of the reflection spectrum $|S^{\hookleftarrow}_{0,0}(\omega)|^2$ vs optomechanical coupling calculated  for $\phi=\pi$ and $\Gamma_0/\Omega=2$. Dashed blue lines label the energies of the two lowest eigenstates of $H_{\rm eff}^{-}$. }
    \label{fig_bright}
\end{figure}

The darkening of the $d$ phonon eigenstates in the \PT-broken regime discussed above should lead to the resonant enhancement of the reflection and transmission as the imaginary part of the denominator in Eq.~\eqref{r_full} goes to zero for the corresponding intermediate eigenstates. In order to demonstrate this effect, we show in Fig.~\ref{fig:T01}(b) by color the absolute squared transmission coefficient with emission of 1 $d$ phonon and zero $s$ phonons  as a function of detuning $\delta = \omega-\omega_x$ and $\eta$.  The calculation demonstrates enhancement of the transmission in  the vicinity of the \PT{} symmetry breaking transition for the first two eigenstates of $H_{\rm eff}^{+}$ (red lines), where one of the states becomes almost dark. 

In the case of $\phi=\pi$ (two $\lambda/2$-spaced qubits), the effect of optomechanical coupling is quite the opposite. In the absence of optomechanical coupling the eigenmodes of $H_{\rm eff}^{\pm}$ have the same real energy, however those of  $H_{\rm eff}^{-}$ are dark mode with zero decay rate, while those of $H_{\rm eff}^{+}$ are characterized by the decay rate $2\Gamma_0$. The optomechanical coupling leads to the brightening of the dark modes and to the emergence of the sharp resonances in the reflection/transmission spectra. We illustrate this mechanism in Fig.~\ref{fig_bright}. Namely we plot  the evolution of the coherent reflectance spectrum $|S^{\hookleftarrow}_{0,0}(\omega)|^2$ with $\eta$. We observe the emergence of the sharp dips in the reflection coefficient at the frequencies corresponding to the eigenergies (labelled with dashed blue lines) as $\eta$ increases which corresponds to the brightening of the dark modes due to the optomechanical interaction.

In summary, we have demonstrated that the ultrastrong regime of optomechanical coupling for array of qubits near a waveguide  can be used to engineer optomechanical \PT{} symmetry, achieve strongly subradiant states and control single-photon transmission spectra. We believe that the full potential of this setup is far from being explored and a plethora of new physical phenomena can be  expected. For instance, the full strength of collective light-matter coupling will be uncovered  only for larger  number of qubits $N>2$, where the collective subradiant states start playing role~\cite{Albrecht2019}. Another degree of freedom will open up when more than one photon excitation is taken into account~\cite{Molmer2019,Poshakinskiy2020arXiv}. The resulting combination of long coherence of mechanical vibrations   and strong quantum nonlinearities  can become beneficial for  emerging quantum technologies.

\acknowledgements
 I.I. acknowledges the support  from the Russian Science Foundation (project 20-12-00224). A.N.P.  was supported by the Russian President Grant MD-243.2020.2. A.V.P was supported by the Russian President Grants MD-243.2020.2 and MK-599.2019.2 and the Foundation ``BASIS''. 
 

\bibliography{OptMech_Quant}
\newpage
\begin{widetext}
\appendix*
\section{Supplemental Material: calculation of the Raman scattering amplitude}
\begin{figure*}[b]
  \includegraphics[width=.99\columnwidth]{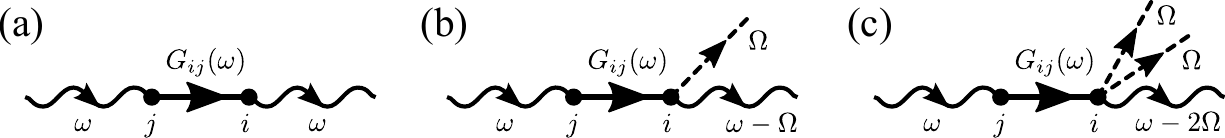}
\caption{Diagrammatic representation of the photon reflection or transmission from a qubit array: (a) coherent process,  (b) Raman process with one phonon emission, (c) Raman process with emission of two phonons.}\label{fig:S1}
\end{figure*}

The amplitude of the single photon transition or reflection from the system is most conveniently calculated using the Green's function technique. The diagrams in Fig.~\ref{fig:S1} show the diagrams corresponding the coherent transmission/reflection [panel (a)] and the Raman scattering processes when one or two vibrational quanta are excited [panels (b) and (c)]. The dot is the photon-phonon-qubit interaction vertex representing Hamiltonian $\hat H_{\rm int}$, Eq.~(2) and  solid line is the Green's function ${ G}_{ij}(\omega)$ that describes light-mediated propagation  of a single exciton from $j$-th to $i$-th qubit dressed by interaction with qubit vibrations. Such Green's function is matrix $N\times N$ of operators acting on vibrational states of all $N$ qubits of the array, that is  defined as ${\bm G}(\omega) = [\omega - {\bm H}_{\rm eff}]^{-1}$, where ${\bm H}_{\rm eff}$ is the matrix of operator Eq.~(3) on the subspace of states with a single qubit excitation,
\begin{align}
H_{{\rm eff},ij} =  \omega_x \delta_{ij}  + \sum_k \Omega a_k^\dag a_k  - \rmi\Gamma_{0}
 \e^{\rmi q |z_i-z_j|}\,\e^{\rmi q u_0\,\text{sign}(z_i-z_j)\, [(a_i + a_i^\dag)- (a_j + a_j^\dag)]} \,.
\end{align}
The amplitude of photon transmission and reflection then reads
\begin{align}
&S^{\rightarrow}_\Phi(\omega) = \delta_{\Phi,0} -\rmi \Gamma_0 \sum_{ij}  \langle \Phi | \e^{-\rmi k[z_i +u_0  (a_i + a_i^\dag)]} G_{ij}  \e^{\rmi k [z_j+ u_0 (a_j + a_j^\dag)]} | 0 \rangle \,,\\
&S^{\hookleftarrow}_\Phi(\omega) =  -\rmi \Gamma_0 \sum_{ij} \langle \Phi | \e^{\rmi k[z_i +u_0  (a_i + a_i^\dag)]} G_{ij}  \e^{\rmi k [z_j+ u_0 (a_j + a_j^\dag)]} | 0 \rangle\:,
\end{align}
where $|0\rangle$ is initial vibrational state, which is supposed to be the ground state, and $|\Phi\rangle$ is the final vibrational state, where some of the qubit vibrations might be excited.

For numerical calculation a coordinate representation of qubit vibrations is more convenient rather than the Fock-state representation. In such representation, the Green's function  $G_{ij}(\omega;\bm x, \bm x')$ satisfies the the equation 
\begin{align}
&\left[ \omega - \omega_x - \frac{\Omega}2 \sum_{k} \left(-2u_0^2\frac{d^2}{dx_k^2} + \frac{x_k^2}{2u_0^2}-1\right)\right] G_{ij}(\omega;\bm x, \bm x')
\\\nonumber &+\rmi \Gamma_{0} \sum_l \e^{\rmi k|z_i- z_l|} \e^{\rmi k \,\text{sign}(z_i- z_l)\, (x_i-x_l)} G_{lj}(\omega;\bm x, \bm x') = \delta_{ij} \delta(\bm x - \bm x') \,.
\end{align}
where $\bm x = (x_1, x_2, \ldots x_N)$ is the vector of coordinates of all qubits. The Green's function is readily expressed as 
\begin{align}\label{eq:SG}
G_{ij}(\omega;\bm x, \bm x') = \sum_{\nu} \frac{\psi^{(\nu)}_i(\bm x) \psi^{(\nu)}_j(\bm x')}{\omega-\omega_x - \epsilon^{(\nu)}}
\end{align}
where $\psi^{(\nu)}_i(x)$ is the set of eigensolutions of the equation system
\begin{align}
  &\frac{\Omega}2 \sum_{k} \left(-2u_0^2\frac{d^2}{dx_k^2} + \frac{x_k^2}{2u_0^2}-1\right) \psi^{(\nu)}_i(\bm x)
\nonumber \\ &-\rmi \Gamma_{0} \sum_k \e^{\rmi q|z_i- z_k|} \e^{\rmi q \,\text{sign}(z_i- z_k)\, (x_i-x_k)} \psi^{(\nu)}_k(\bm x) =
\epsilon^{(\nu)} \psi^{(\nu)}_i(\bm x)
\end{align}
that satisfies the orthogonality condition
$ \sum_i \int  d\bm x \, \psi^{(\nu)}_i(\bm x) \psi^{(\mu)}_i(\bm x)= \delta_{\nu\mu}
$. 
Note that complex conjugation is not involved in this condition, since the effective Hamiltonian is not a Hermitian matrix but a symmetric one. The transmission and reflection coefficients in the coordinate representation read
\begin{align}\label{eq:St}
&S^{\rightarrow}_{\phi}(\omega) = \delta_{\phi, \phi_0}-\rmi\Gamma_{0} \sum_{ij} \int \phi(\bm x) \e^{- \rmi q (z_i+ x_i)} G_{ij}(\omega;\bm x, \bm x')\, \e^{ \rmi q (z_j+ x'_j)} \phi_0(\bm x') \, d\bm x d\bm x' \,,\\\label{eq:Sr}
&S^{\hookleftarrow}_{\phi}(\omega) = -\rmi\Gamma_{0} \sum_{ij} \int \phi(\bm x) \e^{ \rmi q (z_i+ x_i)} G_{ij}(\omega;\bm x, \bm x')\, \e^{ \rmi q (z_j+ x'_j)} \phi_0(\bm x') \, d\bm x d\bm x'\:,
\end{align}
where $\phi_0(\bm x)$ and $\phi(\bm x)$ are the eigenfunctions of the harmonic oscillator corresponding to initial and final vibrational state. 

Applying Eqs.~\eqref{eq:SG},~\eqref{eq:St},~\eqref{eq:Sr} to the case of two-qubit array, as described in the main text, we obtain Eqs.~(7)-(8) of the main tex.  
\end{widetext}

\end{document}